\documentclass[twocolumn,pra]{revtex4}
\usepackage{amssymb}
\usepackage{graphicx}
\usepackage{dcolumn}
\usepackage{bm}
\usepackage{amsmath}
\usepackage{epsfig}

\begin{document}

\title{Entropy exchange and entanglement in the Jaynes-Cummings model}

\author{E. Boukobza}
\author{D.J. Tannor}
\affiliation{Department of Chemical Physics,\\
Weizmann Institute of Science, Rehovot 76100, Israel}

\begin{abstract}

The Jaynes-Cummings model is the simplest fully quantum model that
describes the interaction between light and matter. We extend a
previous analysis by Phoenix and Knight (S. J. D. Phoenix, P. L.
Knight, Annals of Physics {\bf 186}, 381). of the JCM by considering
mixed states of both the light and matter. We present examples of
qualitatively different entropic correlations. In particular, we
explore the regime of entropy \textit{exchange} between light and
matter, i.e. where the rate of change of the two are
anti-correlated. This behavior contrasts with the case of pure
light-matter states in which the rate of change of the two entropies
are positively correlated and in fact identical. We give an
analytical derivation of the anti-correlation phenomenon and discuss
the regime of its validity. Finally, we show a strong correlation
between the region of the Bloch sphere characterized by entropy
exchange and that characterized by minimal entanglement as measured
by the negative eigenvalues of the partially transposed density
matrix.

\end{abstract}

\maketitle

Quantum entropy was first formulated by von Neumann \cite{von
Neumann} as an extension of the Gibbs entropy in classical
statistical mechanics. The foundations of modern information theory
were established by Shannon \cite{Shannon} using a definition of
entropy which is essentially identical to the Gibbs entropy.
Similarly, quantum entropy plays a central role in the theory of
quantum information \cite{Schum} \cite{Schum_West} \cite{Holevo}.
Modern definitions of quantum entropy include those of R\'{e}nyi
\cite{Renyi} and Tsallis \cite{Tsallis}.

In recent years, entropy of quantum systems has been discussed in
the context of entanglement. The Horodecki's \cite{Horod01} derived
R\'{e}nyi entropy inequalities and discussed their implications for
non-locality. They showed that the entropic inequalities are
violated by any pure entangled state. They also argued that by
considering R\'{e}nyi entropy inequalities one can limit
teleportation \cite{teleport} of several states. Cerf \textit{et
al}. \cite{Cerf} gave an interesting interpretation to independent,
correlated, and entangled two qubit systems by considering
conditional and mutual entropies. For example, a maximally entangled
(non-separable) two qubit system possesses negative conditional
entropies and excessive mutual entropy (which is related to
unaccessible information), while a maximally correlated (separable)
two qubit system possesses zero conditional entropies and maximal
classical mutual entropy. Phoenix and Knight \cite{Knight} explored
the dynamics of a single cavity mode in resonance with a single atom
by comparing von Neumann entropy, Shannon entropy, and photon number
variances.

In this paper we focus not on the entropy of the individual
subsystems, but on the entropy \textit{correlations} between
subsystems in a bipartite system.  This question has been almost
completely overlooked in the literature, and for a good reason: for
a bipartite system that starts in a pure state, the partial
entropies of the two subsystems are equal at all times. This result
is a consequence of the equality between the eigenvalues of the
partial density matrices in a pure bipartite system, which in turn
is a consequence of the Schmidt decomposition. In this case the
correlations are trivial. However, there are two qualifications in
obtaining these trivial correlations. First, that the system is
indeed bipartite.  Second, that the joint state of the two
subsystems is pure.  If either of these conditions is not met the
correlation between the entropies of the subsystems is nontrivial.
In particular, in this paper we focus on the question of whether and
under what conditions it is possible to get anti-correlated behavior
of the entropy of the subsystems. This is interesting because it
implies that there is an entropy transfer process going on,
consistent with classical thermodynamic concepts but opposed to the
default result obtained for entangled bipartite systems. In this
paper, for definiteness, we focus on entropy correlations between an
atom and a single quantized cavity mode in the framework of the JCM.
We present qualitatively different entropy correlations between the
atom and the field, and we demonstrate a regime of entropy exchange
between them, both numerically and analytically.

This paper is arranged in the following manner. Section I is a brief
introduction. It is devoted to a description of the density matrix
of bipartite systems, the definition of quantum entropy, and the
issue of separability and entanglement of bipartite systems. In this
section we also review the JCM Hamiltonian in the context of quantum
entropy. Section II is devoted to calculating entropy correlations
for mixed states of light and matter with qualitatively different
behaviors. We give examples of entropy exchange between light and
matter, and explore the parameter range of this regime. In Section
III we give analytical proof of the entropy exchange effect and an
analysis of the regime of its validity. In Section IV we apply two
different entanglement tests to determine whether the atomic-field
system is entangled for all types of entropy correlations. Section V
concludes.

\section{Introduction}

A bipartite system is described by a density matrix of a
$C^{m}\otimes C^{n}$ Hilbert space. The partial density matrix of
one part is obtained by tracing over the other:
\begin{equation}
\rho_{A(B)}=Tr_{B(A)}(\rho_{AB}).
\end{equation}
The entropy of a quantum system is given by the von Neumann entropy
\cite{von Neumann}:
\begin{equation}
S=-k_{B}Tr(\rho\ln\rho).
\end{equation}
The purity of a quantum system is given by $Tr(\rho^{2})$, and it is
bounded: $0<Tr(\rho^{2})\leq1$. Purity is related to the $q=2$
Tsallis entropy \cite{Tsallis}:
\begin{equation}
S_{2}=1-Tr(\rho^{2}).
\end{equation}
Qualitatively, purity oppositely tracks the von Neumann entropy: an
increase in the von Neumann entropy is parallel to a decrease in
purity (or an increase in the Tsallis entropy).

A bipartite quantum system is considered separable if it can be
written as \cite{Werner}:
\begin{equation}
\rho_{AB}=\sum{P^{i}\rho_{A}^{i}\otimes\rho_{B}^{i}},
\label{product}
\end{equation}
where $P^{i}\geq 0$, $\sum{P^{i}}=1$, and $\rho_{A(B)}^{i}$ are
individual partial density matrices. A system that can not be
factored into the form above is said to be entangled. In practice,
determining whether a general $C^{m}\otimes C^{n}$ bipartite system
can be factored into the form above is very hard. Thus, several
tests have been introduced in order to determine whether a system is
entangled. One test originates from quantum information theory, and
it relies on calculating conditional entropies. Conditional entropy
indicates the entropy of one subsystem after measuring the other,
and it is given by:
\begin{equation}
S(A|B)=S_{AB}-S_{B}.
\end{equation}
A bipartite system with at least one degree of freedom having
negative conditional entropy is entangled. Therefore a necessary
condition for separability is that the conditional entropies are
positive. Alternatively, one can calculate the mutual entropy.
Mutual entropy indicates the entropy shared between the two
subsystems, and it is given by:
\begin{equation}
S(A:B)=S_{A}+S_{B}-S_{AB}.
\end{equation}
Mutual entropy is bounded:
\begin{equation}
0\leq S(A:B)\leq 2min[S_{A}, S_{B}].\label{smutlim}
\end{equation}
A bipartite system whose mutual entropy is excessive:
$min[S_{A},S_{B}]<S(A:B)\leq 2min[S_{A},S_{B}]$ is entangled.
Therefore a necessary condition for separability is $0<S(A:B)\leq
min[S_{A},S_{B}]$.

A second powerful test introduced, originally by Peres, \cite{Peres}
relies on partial transposition. Partial transposition is a
blockwise transposition of a matrix and it is given by:
\begin{equation}
\rho_{i\alpha,j\beta}^{T_{2}}\equiv \rho_{i\beta,j\alpha}.
\end{equation}
A system whose partially transposed density matrix is negative is
entangled. Therefore a necessary condition for separability is the
positivity of the partially transposed density matrix (the
Horodecki's \cite{Horod00} showed that it is also a sufficient
condition for separability of $C^{m}\otimes C^{n}$ systems with
$mn\leq 6$). Kraus and coworkers \cite{Kraus} analyzed systems
supported on $C^{2}\otimes C^{N}$. They gave necessary and
sufficient conditions for separability of systems with positive
partially transposed (PPT) density matrices. They showed that if the
rank of density matrix is equal to $N$, the density matrix is
separable. However, other criteria they devised are very complicated
to implement for a generic density matrix.

The JCM is the simplest fully quantum model that describes the
interaction between light and matter. The model consists of a single
quantized two level atom interacting with a single quantized
electromagnetic cavity mode under the rotating wave approximation
(RWA) and the dipole approximation. The resonant JCM Hamiltonian
\cite{JCM} in the interaction representation is given by:
\begin{equation}
H=\hbar\lambda(\sigma_{+}a+\sigma_{-}a^{\dag}).
\end{equation}
For a system that starts in a pure state partial entropies of the
field and the atom are equal at all times for a system that starts
in a pure state \cite{Knight}. As already indicated above, this
result is a consequence of the equality between the eigenvalues of
the partial density matrices in a bipartite system, which in turn is
a consequence of the Schmidt decomposition. Since the partial
entropies fluctuate together in time, their sum does not conserve
the total entropy.
\begin{equation}
S_{a}+S_{f}\geq S_{af}.
\end{equation}
Thus, in this case partial entropies are not an additive (extensive)
quantity. Nevertheless, knowledge of partial entropies gives
information about the dynamics of the total system. The question
that we address here is whether we can find states of light and
matter for which the partial entropies are quasi-additive. Mixed
states of light and matter can be natural candidates in this
respect.

\section{Results}

We consider various combinations of initial pure and mixed states of
light and matter. The mixed state for the electromagnetic field is
chosen to be a Planck distribution for a single cavity mode:
$\rho_{f}(0)=\sum_{n=0}^{\infty}{P_{n}{|n\rangle\langle n|}},\mbox{
}P_{n}=\frac{\bar{n}^{n}} {(\bar{n}+1)^{n+1}}$, where $\bar{n}$ is
the average number of photons in the cavity. The mixed state for the
atom has the following general form:
$\rho_{a}(0)=P_{e}|e\rangle\langle e| +(1-P_{e})|g\rangle\langle
g|,\mbox{ }0<P_{e}<1$. When the cavity mode is in exact resonance
with the atomic transition, one can derive an analytic expression
for the propagator and obtain analytic expressions for the full and
partial density matrices at any time. The general solution for the
full density matrix has the following form:
\begin{equation}
\rho_{af}(t)=e^{-iHt}\rho_{af}(0)e^{iHt},
\end{equation}
where $H$ is the interaction Hamiltonian mentioned above. By
expanding $U(t)=e^{-iHt}$ in a Taylor series one obtains the
analytic form of the propagator:
\begin{equation}
U(t)=\begin{pmatrix}
  \cos(\lambda t\sqrt{aa^{\dag}}) & -i\frac{\sin(\lambda t\sqrt{aa^{\dag}})}{\sqrt{aa^{\dag}}}a \nonumber\\
  -i\frac{\sin(\lambda t\sqrt{a^{\dag}a})}{\sqrt{a^{\dag}a}}a^{\dag} & \cos(\lambda
  t\sqrt{a^{\dag}a})
\end{pmatrix}
\end{equation}
To simulate the evolution of the entropic quantities, we truncated
the set of Fock states that compose the Planck distribution at some
$n_{f},$ where $\sum_{n=0}^{n_{f}}{P_{n}}\cong 1$. The accuracy of
the results we obtained was tested by adding more Fock states to the
Planck distribution to see if the simulated values changed.

\begin{figure}[htb]
\begin{center}
\includegraphics[width=8cm]{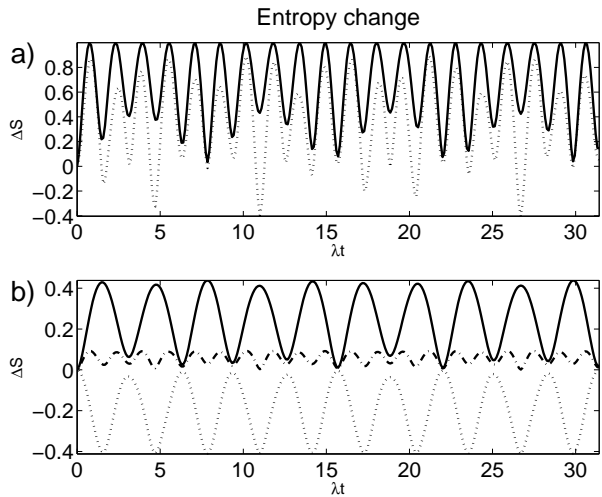}
\end{center}
\caption{Partial entropy change plots $\Delta S=S-S(0)$ (field
dotted line, atomic solid line). Top: atom initially in the excited
state and a weakly excited thermal field, $\rho(0)=(|e\rangle\langle
e|)_{a}\otimes (\sum_{n=0}^{\infty}{P_{n}{|n\rangle\langle n|}};
\bar{n}=0.1)_{f}$. Bottom: atom initially in the ground state and a
weakly excited thermal field , $\rho(0)=(|g\rangle\langle
g|)_{a}\otimes (\sum_{n=0}^{\infty}{P_{n}{|n\rangle\langle
n|}})_{f}; \bar{n}=0.1)_{f}$. The dash-dot curve represents the sum
of atomic and field partial entropy changes, and it is
quasi-conserved.} \label{fig1}
\end{figure}
In Fig. \ref{fig1} the changes in atomic and field partial entropies
($S(t)-S(0)$) vs. the dimensionless quantity $\lambda t$ are
plotted. In Fig.1a the atom is initially excited and the field is in
a weakly excited thermal state ($\bar{n}=1$). This case is similar
to a situation where both the field and atom are initially in a pure
state, because both partial entropies rise and lower together
(although they are not equal). In Fig. 1b the atom is initially in
the ground state and the field is in a weakly excited thermal state.
It is clear that there is entropy exchange between the atom and
field (although the exchange is not complete). The sum of the field
and atomic entropy changes (green curve) is quasi-conserved as seen
by its amplitude of fluctuations, which is significantly smaller
than each of the partial entropy changes.

When the field is excited more, i.e. $\bar{n}=1$, or $\bar{n}=10$,
no substantial entropy exchange between the atom and field occurs.
In particular, when the field is highly excited ($\bar{n}\geq 10$)
and the atom is close to a pure state, there is a sharp and rapid
collapse of the atomic purity (sharp and rapid rise in the von
Neumann entropy) with no substantial revival.

In order to determine which initial atomic states can exchange
entropy with a weakly excited thermal field we considered different
initial atomic states by discretizing the Bloch sphere with a
longitudinal angle ($-\frac{\pi}{2}\leq\theta\leq\frac{\pi}{2}$, for
positive $\theta$ the atom is more excited) and a Bloch vector
length ($0<r\leq1$), disregarding the azimuthal angle ($\phi$). We
introduce a time averaged entropy exchange parameter defined by:
\begin{equation}
P=\overline{(\frac{\Delta S_{a(f)}}{\Delta S_{f(a)}})},
\end{equation}
where $\Delta S_{a(f)}=S_{a(f)}(t_{j})-S_{a(f)}(t_{j-1})$, and in
the numerator the smaller of the two instantaneous partial entropy
changes was substituted. The time averaged entropy exchange
parameter is suitable for quantifying the extent of entropy flow
between a weakly excited thermal field and an atom due to the
oscillatory nature of the partial entropies. Furthermore, the
entropy exchange parameter is bounded: $-1\leq P \leq 1$; when it
tends to $-1$ there is high degree of entropy exchange, and when it
tends to $+1$ the two partial entropies rise and lower together.

\begin{figure}[htb]
\begin{center}
\includegraphics[width=8cm]{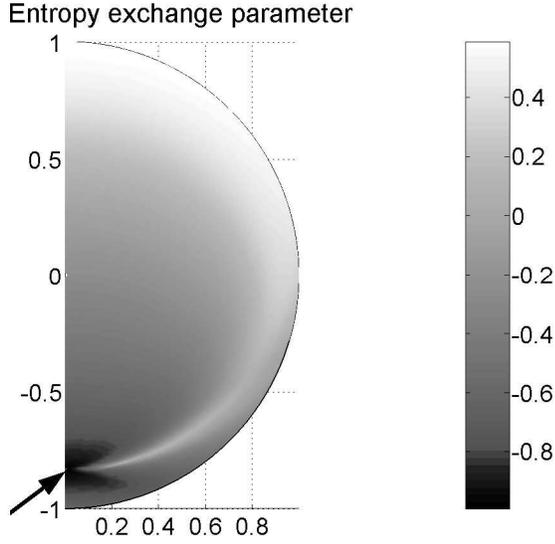}
\end{center}
\caption{Entropy exchange parameter for a weakly excited thermal
field and various initial atomic density matrices, characterized by
the longitudinal angle $\theta$ and the Bloch vector length $r$. The
dark region in parameter space is characterized by significant
entropy exchange ($P<-0.8$). Note that this region is centered
around a 'fixed' point where the partial density matrices are
stationary (black arrow; see text).} \label{fig2}
\end{figure}

\begin{figure}[htb]
\begin{center}
\includegraphics[width=8cm]{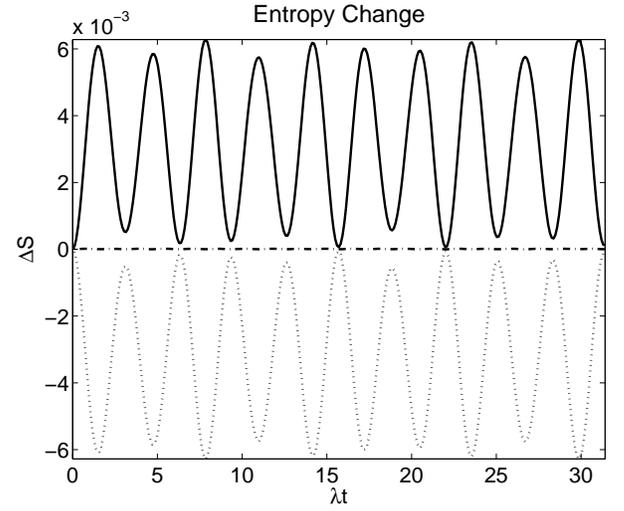}
\end{center}
\caption{Partial entropy change (field dotted line, atomic solid
line, atomic+field dash dot curve) for a weakly excited thermal
field ($\bar{n}=0.1$) and atom initially in the state $\rho_{a}
(r=\sqrt{\frac{7}{10}};\theta=-\frac{\pi}{2})$. Entropy exchange
between the atom and field is almost complete ($P\approxeq 0.99$).}
\label{fig3}
\end{figure}

In Fig. \ref{fig2} the entropy exchange parameter is plotted for a
weakly excited thermal field and various initial atomic states. The
dark blue region in parameter space is characterized by significant
entropy exchange ($P<-0.8$); and in this region the atom is mostly
in the ground state (ground state probability is more than $80\%$).
Thus entropy exchange occurs most effectively when the atom is
initially close to the ground state. An example for an almost
complete entropy exchange is given in Fig. \ref{fig3}, where the
changes in atomic and field partial entropies are shown for an atom
initially in the state $\rho_{a}(r=0.7, \theta=-\frac{\pi}{2})$
($P\approxeq 0.99$). In this case the sum of the two partial
entropies is almost completely conserved as seen by the amplitude of
fluctuation of the sum of the partial entropy changes, which is two
orders of magnitude smaller than each of the partial entropy
changes.

\section{Analytical derivation of the entropy exchange effect}
We now provide an analytical derivation for the entropy exchange
effect. We first note that contours of significant entropy exchange
($P<-0.8$, Fig. \ref{fig2}) center around a point on the central
vertical axis of the Bloch sphere. This is a 'fixed' point where the
partial density matrices are stationary, and it corresponds to a
situation where the field temperature is identical to the atomic
Boltzman temperature as we shall now show. Consider a situation
where:
\begin{equation}
\frac{P_{e}}{P_{g}}=\frac{P_{n+1}}{P_{n}}=\frac{\bar{n}}{\bar{n}+1}=e^{-\frac{\hbar\omega}{k_{B}T}},\label{ratios}
\end{equation}
where $\omega$ is the transition frequency. The full density matrix
where both the atom and field are in mixed states (with no initial
coherence) is given by:
\begin{equation}
\rho_{af}(t)=U\begin{pmatrix}
P_{e}\otimes\mathbf{P_{n}} & 0 \nonumber\\
0 & P_{g}\otimes\mathbf{P_{n}}
\end{pmatrix}U^{\dag},
\end{equation}
where $(\mathbf{P_{n}})_{mn}=\delta_{mn}P_{n}$. The partial density
matrices are in diagonal form:
\begin{eqnarray}
\rho_{a}(t)&=&\begin{pmatrix}
  A(t) & 0 \nonumber\\
  0 & B(t)
\end{pmatrix}\\
\rho_{f}^{nn}(t)&=&C(t)+D(t),\label{mixeda}
\end{eqnarray}
where
\begin{eqnarray}
A(t)&=&P_{e}\sum{P_{n}\cos^{2}(\alpha_{n}t)}+P_{g}\sum{P_{n+1}\sin^{2}(\beta_{n+1}t)},\nonumber\\
B(t)&=&P_{e}\sum{P_{n-1}\sin^{2}(\alpha_{n-1}t)}+P_{g}\sum{P_{n}\cos^{2}(\beta_{n}t)},\nonumber\\
C(t)&=&P_{e}(P_{n}\cos^{2}(\alpha_{n}t)+P_{n-1}\sin^{2}(\alpha_{n-1}t)),\nonumber\\
D(t)&=&P_{g}(P_{n+1}\sin^{2}(\beta_{n+1}t)+P_{n}\cos^{2}(\beta_{n}t)),\nonumber\\
\alpha_{n-1}&=&\beta_{n}=\lambda\sqrt{n}.
\end{eqnarray}
By substituting equation \ref{ratios} into equation \ref{mixeda} one
obtains:
\begin{eqnarray}
\rho_{a}(t)&=&\rho_{a}(0)=\begin{pmatrix}
  P_{e} & 0 \nonumber\\
  0 & P_{g}
\end{pmatrix}\\
\rho_{f}^{nn}(t)&=&\rho_{f}^{nn}(0)=P_{n}.
\end{eqnarray}
For example for a weakly excited thermal field with $\langle
\bar{n}\rangle=0.1$, this point is located at $r=\frac{5}{6};
\theta=-\frac{\pi}{2}$ (as indicated by the black arrow in Fig.
\ref{fig2}).

We now proceed to analyze the vicinity around the fixed point, where
substantial entropy exchange occurs.  For a weakly excited thermal
field the atomic fixed point is close to the ground state (see Fig.
\ref{fig2}). Therefore, we consider an initial state where an atom
in the ground state interacts with a thermal cavity mode:
$\rho_{af}(0)=(|g\rangle\langle g|)_{a}\otimes
(\sum_{n=0}^{\infty}{P_{n}{|n\rangle\langle n|}})_{f}$. In this case
the full density matrix is given by:
\begin{equation}
\rho_{af}(t)=\begin{pmatrix}
S_{1}\rho_{f}(0)S_{1}^{\dag} & -iS_{1}\rho_{f}(0)C_{1}^{\dag} \nonumber\\
iC_{1}\rho_{f}S_{1}^{\dag} & C_{1}\rho_{f}(0)C_{1}^{\dag}
\end{pmatrix}
\end{equation}
where $C_{1}=\cos(\lambda t\sqrt{a^{\dag}a})$, and
$S_{1}=\frac{\sin(\lambda t\sqrt{aa^{\dag}}}{\sqrt{aa^{\dag}}}a$.
The partial density matrices are in diagonal form:
\begin{eqnarray}
\rho_{a}(t)&=&\begin{pmatrix}
  \sum{P_{n+1}\sin^{2}(\beta_{n+1}t)} & 0 \nonumber\\
  0 & \sum{P_{n}\cos^{2}(\beta_{n}t)}
\end{pmatrix}\\
\rho_{f}^{nn}(t)&=&P_{n}\cos^{2}(\beta_{n}t)+P_{n+1}\sin^{2}(\beta_{n+1}t).
\end{eqnarray}
Since the purity is qualitatively similar to the von Neumann entropy
we can analyze the time dependence of the partial purities. The time
derivatives of the partial purities are given by:
\begin{eqnarray}
\frac{d(Tr(\rho_{a}^{2}))}{dt}&\!\!\!\!\!\!=&\!\!\!\!\!\!2\sum P_{n+1}\sin^{2}(\beta_{n+1}t)\sum P_{n+1}\beta_{n+1}\sin(2\beta_{n+1}t)\nonumber\\
&-&\!\!2\sum P_{n}\cos^{2}(\beta_{n}t)\sum
P_{n}\beta_{n}\sin(2\beta_{n}t)\nonumber\\
\frac{d(Tr(\rho_{f}^{2}))}{dt}&=&\!\!\!2\sum(P_{n+1}\beta_{n+1}\sin(2\beta_{n+1}t)-P_{n}\beta_{n}\sin(2\beta_{n}t))\nonumber\\
&\cdot&\!\!(P_{n}\cos^{2}(\beta_{n}t)+P_{n+1}\sin^{2}(\beta_{n+1}t)).
\label{dSRga}
\end{eqnarray}
One can rearrange the time derivatives in equation \ref{dSRga} in
terms of $P_{i}P_{j}$ multiplying sums of oscillatory functions. For
a weakly excited thermal cavity ($\bar{n}=0.1$) the $P_{i}P_{j}$
decrease by approximately an order of magnitude for every increment
in either $i$ or $j$. Approximating the derivatives of the partial
purities by the first non vanishing term ($P_{0}P_{1}=0.0751$) one
obtains:
\begin{eqnarray}
\frac{d(Tr(\rho_{a}^{2}))}{dt}&\approx&-2P_{0}P_{1}\beta_{1}\sin(2\beta_{1}t)\nonumber\\
\frac{d(Tr(\rho_{f}^{2}))}{dt}&\approx&2P_{0}P_{1}\beta_{1}\sin(2\beta_{1}t).\label{dSRgappa}
\end{eqnarray}
The above approximation shows explicitly that there is entropy
exchange between a weakly excited thermal cavity mode and an atom
initially in the ground state. The leading terms in the time
derivatives of the partial purities have the same functional
dependence on time but with opposite signs.

Consider now an initial state of an excited atom interacting with a
thermal cavity mode: $\rho_{af}(0)=(|e\rangle\langle e|)_{a}\otimes
(\sum_{n=0}^{\infty}{P_{n}{|n\rangle\langle n|}})_{f}$. In this case
the full density matrix is given by:
\begin{equation}
\rho_{af}(t)=\begin{pmatrix}
C_{2}\rho_{f}(0)C_{2}^{\dag} & iC_{2}\rho_{f}(0)S_{2}^{\dag} \nonumber\\
-iS_{2}\rho_{f}C_{2}^{\dag} & S_{2}\rho_{f}(0)S_{2}^{\dag}
\end{pmatrix}
\end{equation}
where $C_{2}=\cos(\lambda t\sqrt{aa^{\dag}})$, and
$S_{2}=\frac{\sin(\lambda
t\sqrt{a^{\dag}a})}{\sqrt{a^{\dag}a}}a^{\dag}$. Again the partial
density matrices are in diagonal form:
\begin{eqnarray}
\rho_{a}(t)&=&\begin{pmatrix}
  \sum{P_{n}\cos^{2}(\alpha_{n}t)} & 0 \nonumber\\
  0 & \sum{P_{n-1}\sin^{2}(\alpha_{n-1}t)}
\end{pmatrix}\\
\rho_{f}^{nn}(t)&=&P_{n}\cos^{2}(\alpha_{n}t)+P_{n-1}\sin^{2}(\alpha_{n-1}t).
\end{eqnarray}
The time derivatives of the partial purities are given by:
\begin{eqnarray}
\frac{d(Tr(\rho_{a}^{2}))}{dt}&=&-2\sum
P_{n}\alpha_{n}\sin(2\alpha_{n}t)\sum
P_{n}\cos^{2}(\alpha_{n}t)\nonumber\\
&\!\!\!\!\!\!\!\!+&\!\!\!\!\!\!\!\!2\sum
P_{n-1}\sin^{2}(\alpha_{n-1}t)\sum
P_{n-1}\alpha_{n-1}\sin(2\alpha_{n-1}t)\nonumber\\
\frac{d(Tr(\rho_{f}^{2}))}{dt}&\!\!\!\!=&\!\!\!\!2\sum(P_{n-1}\alpha_{n-1}\sin(2\alpha_{n-1}t)-P_{n}\alpha_{n}\sin(2\alpha_{n}t))\nonumber\\
&\cdot&(P_{n}\cos^{2}(\alpha_{n}t)+P_{n-1}\sin^{2}(\alpha_{n-1}t)).
\label{dSRgb}
\end{eqnarray}
Rearranging the time derivatives in equation \ref{dSRgb} in terms of
$P_{i}P_{j}$, as in the case when the atom was initially in the
ground state, and approximating them by the first non vanishing term
($P_{0}^{2}=0.8264$) one obtains:
\begin{eqnarray}
\frac{d(Tr(\rho_{a}^{2}))}{dt}&\approx&-P_{0}^{2}\alpha_{0}\sin(4\alpha_{0}t)\nonumber\\
\frac{d(Tr(\rho_{f}^{2}))}{dt}&\approx&-P_{0}^{2}\alpha_{0}\sin(4\alpha_{0}t).\label{dSReappa}
\end{eqnarray}
The above approximation shows that in this case the partial purities
oscillate together in time. The leading terms in the time
derivatives of the partial purities are identical.

When the cavity mode is excited thermally with more photons, i.e.
$\bar{n}\geq1$, the terms in the time derivatives of the partial
purities rearranged according to $P_{i}P_{j}$ no longer decrease by
an order of magnitude as $i, j$ increase. Now many terms contribute
substantially to the partial entropy behavior; since these terms are
not identical in both degrees of freedom, observing entropy exchange
becomes increasingly less probable as $\bar{n}$ increases.

\section{Entropy correlations and entanglement}

We consider now the relationship between entropy correlations
developed between the atom and field discussed in previous sections,
and entanglement of the various atomic-field states. One would
expect that as $S_{A}+S_{B}\rightarrow S_{AB}$ the system becomes
separable. This is motivated by the following formula from
statistical mechanics \cite{Reif}:
\begin{equation}
S_{AB}=S_{A}+S_{B}+k_{B}\sum_{i,j}{\ln\frac{P_{A}^{i}P_{B}^{j}}{P_{AB}^{ij}}}.
\end{equation}
If $S_{AB}=S_{A}+S_{B}$, then $P_{AB}^{ij}=P_{A}^{i}P_{B}^{j}$
(since the sum over the natural logarithm terms has to be zero).
This means that the two subsystems are independent (if
$S_{A}+S_{B}\cong S_{AB}$ then the two subsystems are weakly
classically correlated). In regions of the Bloch sphere where the
entropy exchange is almost complete ($\Delta S_{a}=\simeq -\Delta
S_{f}$):
\begin{equation}
S_{a}+S_{f}\simeq S_{a}(0)+S_{f}(0)=S_{af}(0)=S_{af}.
\end{equation}
The last two equalities hold since the combined atomic-field system
begins to evolve from a separable state, and since the evolution is
unitary, respectively. Therefore, we would expect that in regions of
the Bloch sphere where the entropy exchange is almost complete,
there should be minimal entanglement. We now proceed with comparing
our measure for entropy exchange with various entanglement measures.

\begin{figure}[htb]
\begin{center}
\includegraphics[width=8cm]{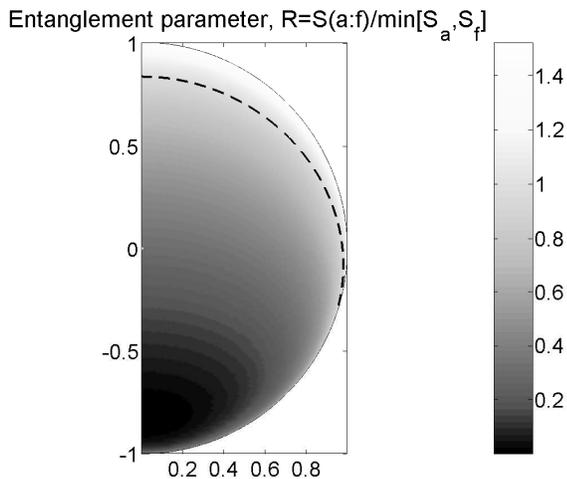}
\end{center}
\caption{Time averaged entanglement parameter based on the ratio
between mutual entropy and partial entropies for different initial
atomic states coupled to a weakly excited thermal field
($\bar{n}=0.1$). The region in parameter space where $\bar{R}\leq 1$
(dotted line) encompasses the entire region of entropy exchange ($P
< -0.8$), but in addition includes significant area where there is
no entropy exchange (even regions with $P>0$; compare with fig.
\ref{fig2})} \label{fig4}
\end{figure}

As discussed in the Introduction, bipartite systems with excessive
mutual entropy are entangled. We introduce an entanglement parameter
$R$ based on the ratio between mutual entropy and partial entropies:
\begin{equation}
R=\frac{S(a:f)}{min[S_{a},S_{f}]}\label{R}
\end{equation}
It follows from equation \ref{smutlim} that $0\leq R\leq 2$, with
$1<R \leq 2$ being a sufficient condition for entanglement and
conversely $0 \leq R \leq 1$ being a necessary condition for
separability. In fig. \ref{fig4} the time averaged entanglement
parameter, $\bar{R}$, is plotted for different initial atomic states
coupled to a weakly excited thermal field. The region in parameter
space with $\bar{R}\leq 1$ (bounded by a dashed line) indicates that
the {\em necessary} condition for separability is fulfilled for most
of the evolution time. This region is large, occupying about 75\% of
the interior of the Bloch sphere. It encompasses the entire region
of entropy exchange ($P < -0.8$), but in addition includes
significant area where there is no entropy exchange (even regions
with $P>0$). This is consistent with the observation that
$0<\bar{R}\leq 1$ is a {\em necessary} but not a {\em sufficient}\
condition for separability.  Although the region defined by this
necessary condition is too broad to be informative, note that in
cases where there is substantial entropy exchange as seen in fig.
\ref{fig2}, $\bar{R}$ approaches zero in fig. \ref{fig4}.

\begin{figure}[htb]
\begin{center}
\includegraphics[width=8cm]{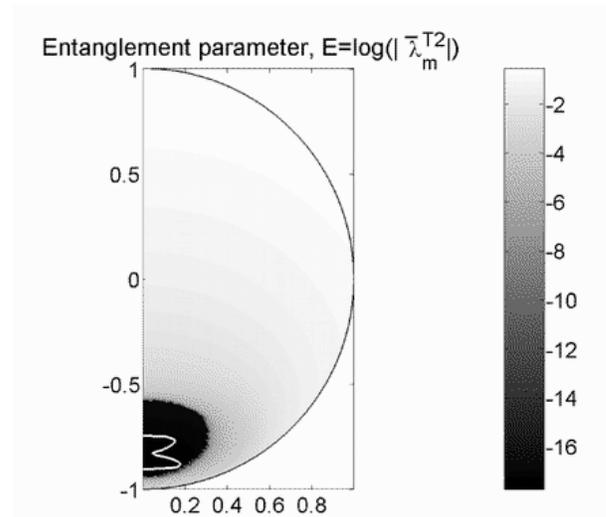}
\end{center}
\caption{Entanglement parameter based on the negativity of the
eigenvalues of the partially transposed density matrix for a weakly
excited thermal field ($\bar{n}\approx 0.1$) and various atomic
states. The region of atomic parameter space with substantial
entropy exchange ($P<-0.8$, bounded by a solid white line) falls
within the region with minimal entanglement parameter (dark), which
is dominated by anti-correlated partial entropies.} \label{fig5}
\end{figure}

We consider now the PPT test, where the existence of even a single
negative eigenvalue of the partially transposed density matrix is a
{\em sufficient} criterion for entanglement. Since the PPT test can
only be applied for finite dimensional density matrices, in order to
apply it to the Jaynes-Cummings model we need to truncate the
infinite set of Fock states. We do this according to the following
procedure:
\begin{eqnarray}
\rho_{f}(0)&=&\sum_{n=0}^{\infty} P_n |n\rangle\langle n| \\
&\!\!\!\!\rightarrow&\!\!\!\!
\sum_{n=0}^{n_{f}}{P_{n}|n\rangle\langle
n|}+(1-\!\!\sum_{n=0}^{n_{f}}{P_{n}})|n_{f+1}\rangle\langle
n_{f+1}|;
\end{eqnarray}
in words, all residual probability from the truncation is placed in
the $n_{f}+1$st state. The criterion for truncating the infinite set
of Fock states is that the partial entropies do not change by more
than $10^{-14}$. We find that no matter what the initial state of
the atom is, there is always at least one negative eigenvalue of the
partially transposed density matrix. However, further investigation
shows that this result is in essence due to the truncation of the
infinite Fock basis to a finite size.  Specifically, we find that in
all regions of the Bloch sphere there is always one negative
eigenvalue of extremely small magnitude whose size is on the order
of the probabilities tail of the infinite Fock basis that gets
truncated.  Since our primary interest here is in the original and
not in the truncated Jaynes-Cummings model, we are inclined to
ignore this negative eigenvalue of extremely small magnitude and to
interpret the test of partial transposition in terms of the
remaining negative eigenvalues.  The remaining negative eigenvalues
are all of sizable magnitude and robust with respect to the size of
the Fock basis.

With that introduction, we define $\lambda_m^{T_2}$ as the largest
(in absolute value) negative eigenvalue of the partially transposed
density matrix, $\rho^{T_{2}}$.  We then introduce as an
entanglement measure
\begin{equation}
E=\log(|\bar{\lambda}^{T_{2}}_{m}|),
\end{equation}
where the overbar signifies the time average.  In figure \ref{fig5}
we plot the entanglement parameter, $E$, for different initial
states for $n_{f}=13$. Clearly,  there are two well-defined regions
in atomic parameter space. The dark region corresponds to a single
negative eigenvalue of $\rho^{T_{2}}$ on the order of
$10^{-16}-10^{-18}$. This region is characterized by marginal
entanglement, corresponding in all likelihood to completely
separable evolution in the untruncated Jaynes-Cummings model. Notice
that the region of atomic parameter space with substantial entropy
exchange ($P<-0.8$ in fig. \ref{fig2}, shown as a solid white line
in fig. \ref{fig5}) falls precisely within the region with minimal
entanglement parameter, $E$. The lighter shade region in fig.
\ref{fig5} is characterized by two or more negative eigenvalues of
$\rho^{T_{2}}$. We believe that this region corresponds to
significant entanglement in the original, untruncated
Jaynes-Cummings model. This region is clearly orthogonal to that of
the region of entropy exchange.

\section{Conclusion}

We have explored entropy correlations between a quantized cavity
mode and a single atom in the framework of the JCM by considering
both pure and mixed atomic and field states. In particular, we
explored the regime of entropy \textit{exchange} between light and
matter. We presented two qualitatively different entropy
correlations. The first type of correlation is a case where both the
atomic and field partial entropies fluctuate together in time. This
is reminiscent of the case where both the atom and field start out
in a pure state, and consequently their partial entropies are
identical at all times. The second type of correlation is a case
where the atomic and field partial entropies are anti-correlated.
This implies that there is entropy exchange between the atom and the
field. Since substantial entropy exchange occurs when the field is
in a weakly excited thermal state ($\bar{n}=0.1$) we introduced an
entropy exchange parameter in order to determine which initial
atomic states can efficiently exchange entropy with the field.
Substantial entropy exchange occurs when the atom is initially close
to the ground state. We showed that contours of substantial entropy
exchange center around a stationary point in the Bloch sphere. This
point corresponds to a situation where the initial field and atomic
(Boltzman) temperature exactly match. By analyzing the partial
purities we derived an analytic approximation for the entropy
exchange phenomenon. The change in the atomic and field partial
purities have the same functional dependence on time but with
opposite signs.

It is natural to ask if there is any connection between the entropy
exchange that we observe in this paper and a thermodynamic heat
exchange process.  Although we cannot rule out this possibility, we
believe that the connection is unlikely.  In classical
thermodynamics, heat exchange is related to the transfer of energy
that is in some sense degraded.  In our case, because the entropy
exchange is periodic and the system is small, there is no reason to
believe a priori that the energy that is exchanged is unrecoverable
for pure work. Nevertheless, if one examines the ratio
$\frac{dE/dt}{dS/dt}$, where $dE/dt$ and $dS/dt$ are the
instantaneous rates of energy and entropy change of each subsystem
respectively, there is some overall correlation with the initial
temperature $T$ of the field, in agreement with the classical result
$\frac{dE}{dS}=T$.

A well-established measure of entanglement is the test of negative
eigenvalues of the partial transpose of the density matrix (PPT
test).  Unfortunately, the PPT test cannot be applied directly to
the Jaynes-Cummings model with its formally infinite Hilbert space
for the field.  We therefore applied the PPT test to a model in
which the infinite Fock basis was truncated.  We found that there is
always at least one negative eigenvalue for the partially transposed
density matrix.  However, further investigation showed that this
result is due to the truncation of the Fock basis.  Since our
primary interest here is in the original and not in the truncated
Jaynes-Cummings model, we essentially discard this negative
eigenvalue of extremely small magnitude and interpret the test of
partial transposition in terms of only the remaining negative
eigenvalues.  We find that the region of the Bloch sphere in which
there are are no additional negative eigenvalues maps very closely
onto the region where where there is substantial entropy exchange
between atom and field.  This result is intuitively appealing,
showing a strong negative correlation between entropy exchange and
entanglement.

\section*{Acknowledgments}
We thank Prof. Sam Braunstein, and Prof. Mikhail Lukin for helpful
comments.

\bibliographystyle{phaip}

\end{document}